\documentclass[review]{elsarticle}

\usepackage{hyperref}

\usepackage{graphicx}
\usepackage[all]{hypcap} \usepackage{listings}

\usepackage{color}
\usepackage{xspace}
\usepackage{macros}
\usepackage{aas_journals}

\journal{Astronomy and Computing}

\begin{document}
\def\phxO{{\tt PHOENIX/1D}\xspace}
\def\phxT{{\tt PHOENIX/3D}\xspace}
\def\phx{{\tt PHOENIX}\xspace}

\lstset{language=Fortran}

\begin{frontmatter}

\title{A 3D radiative transfer framework: XII. Many-core, vector and GPU methods}

\author{Peter H. Hauschildt}
\address{Hamburger Sternwarte, Gojenbergsweg 112, 21029 Hamburg, Germany}
\ead{yeti@hs.uni-hamburg.de}

\author{E.~Baron}
\address[mymainaddress]{Homer L.~Dodge Dept.~of Physics and Astronomy, University of
Oklahoma, 440 W.  Brooks, Rm 100, Norman, OK 73019 USA}
\address[mysecondaryaddress]{Department of Physics and Astronomy, Aarhus University, Ny Munkegade 120, DK-8000 Aarhus C, Denmark}
\ead{baron@ou.edu}

\begin{abstract}
3D detailed radiative transfer is computationally taxing, since the
solution of the radiative transfer equation involves traversing the six
dimensional phase space of the 3D domain. With modern supercomputers the
hardware available for wallclock speedup is rapidly changing, mostly in response
to requirements to minimize the cost of electrical power.
Given the variety of modern computing architectures, we aim to develop
and adapt algorithms for different computing architectures to improve
performance on a wide variety of platforms. We implemented the main time
consuming kernels for solving 3D radiative transfer problems for vastly
different computing architectures using MPI, OpenMP, OpenACC and vector
algorithms. Adapted algorithms lead to massively improved speed for all
architectures, making extremely large  model calculations easily
feasible. These calculations would have previously been considered
impossible or prohibitively expensive. Efficient use of modern computing
devices is entirely feasible, but unfortunately requires the
implementation of specialized algorithms for them.

\end{abstract}


\end{frontmatter}

\section{Introduction}

In a series of papers, we
have described a framework for solving the radiative transfer equation
in 3D systems (3DRT), including a detailed treatment of scattering in continua
and lines with a non local operator splitting method and its use in the
general model atmosphere package \phx \citep[][hereafter: Papers I--XI]{3drt_paper1,
3drt_paper2, 3drt_paper3, 3drt_paper4,
3drt_paper5,3drt_paper6,3drt_paper7,3drt_paper8,3drt_paper9,3drt_paper10,3drt_paper11}.

We give a short summary of the problem and the numerical approach in the next section.

In
typical non-local thermodynamic equilibrium (NLTE) \phxT\ applications, 
the 3DRT uses about 75\% of the overall compute time, that is, reducing the 
time consumed by the 3DRT module will significantly reduce the 
overall simulation time. This is an advantage of our approach since
only a small subset of code modules require specialized directives and
code modifications to match the targeted hardware and compiler.

In Paper VIII \citep[][]{3drt_paper8} we showed that specialized codes for GPUs can
result in significantly improved performance compared to standard CPUs of the
era. With the availability of new many-core CPUs (e.g., Intel Phi), vector
CPUs (i.e., NEC Aurora TSUBASA) and GPUs, the need for  algorithms
specially adapted to the hardware  in order to maximize 
performance on such systems becomes apparent. This becomes more urgent
due the fact that  modern 
supercomputers are being built using many-core CPUs or as hybrid CPU/GPU
systems. While, it is likely that efficient use of such systems will require
specialized algorithms,  it is not economical to design special codes
for each system or to use vendor-locked programming models as
individual computing system lifetimes 
(5 years) are typically much shorter than code lifetimes (20+ years).

For these reasons, we investigate here how 3D radiative transfer calculations
can be accelerated using algorithms  adapted to many-core and vector
CPUs as well as to GPUs. As codes will be used potentially for decades, it is very
important to adhere to general standards as closely as possibly in order to
retain source code compatibility for long time scales.  Thus, we  will use Fortran 2008
\citep[][]{F2008} as the base programming language and will use MPI \citep[][]{MPI},
OpenMP \citep[][]{OpenMP} and OpenACC \citep[][]{OpenACC} as additional directive
based performance enhancers. In addition, directive based statements for
specific systems, e.g., to vectorize loops, are considered as long as they do
not interfere with portability to other systems. 

In the following, we consider two test cases that represent typical 3DRT usage
patterns of \phxT: A Cartesian box with periodic boundary conditions in the
horizontal (x,y) plane and a spherical coordinate system case used in
irradiation models in pre-CVs or exoplanets. These general setups cover the
vast majority of current and planned \phxT\ calculations, accelerating them can
save millions of CPU-core-hours and related energy expenditures.  In
addition, faster 
calculations can enable the use of smaller supercomputers so that not only Tier
0 systems, but also Tier 1 or 2 systems can be used for a given model. This can
drastically reduce the turnaround time for a model and in turn, free up Tier 0
systems for larger model runs.

\section{3D radiative transfer framework}

We\footnote[1]{This section is adapted from \cite{3drt_paper1}} use a Cartesian or spherical coordinates grid of non-equal sized volume
cells (voxels) for the following discussion.  The values of physical
quantities, such as temperatures, opacities and mean intensities, are averages
over a voxel, which, therefore, also fixes the local physical resolution of the
grid. In the following we will specify the size of the voxel grid by the number
of voxels along each positive axis, e.g., $n_x = n_y = n_z = 32$ specifies a
voxel grid from voxel coordinates $(-32,-32,-32)$ to $(32,32,32)$ for a total
of  $(2*32+1)^3 = 274625$ voxels, $65$ along each axis. 

The 3DRT framework is typically applied to 
optically thick environments with a significant scattering contribution,
e.g., modeling the light reflected by an extrasolar giant planet close to 
its parent star. Therefore, not only a formal solution (see below) of the radiative transfer equation is required, but the 
full solution of the 3D radiative transfer equation with scattering. 
The fundamental numerical method used in the 3DRT framework is 
an operator splitting approach. Operator splitting works best if 
a non-local $\lstar$ operator is used in the calculations \cite[e.g.,][]{aliperf},
therefore, we exclusively use a non-local $\lstar$ method.

{
\subsection{Radiative transfer equation}

For simplicity, we consider here only the  static, steady-state radiative transfer equation in
3D (see \cite[][]{3drt_paper9,3drt_paper10} for the non-static case)  written
as
\begin{equation}
\hat{\vec n} \cdot \nabla
I(\nu,\vec x,\hat{\vec n}) = \eta(\nu,\vec x) - \chi(\nu,\vec x)I(\nu,\vec
x,\hat{\vec n})
\end{equation}
where $I(\nu,\vec x,\hat{\vec n})$ is the specific intensity at frequency
$\nu$, position $\vec x$, in the direction $\hat {\vec n}$, $\eta(\nu,\vec x)$
is the emissivity at frequency $\nu$ and position $\vec x$, and $\chi(\nu,\vec
x)$ is the total extinction at frequency $\nu$ and position $\vec x$. The
source function $S = {\eta}/{\chi}$. Mathematically, this is a linear first
order partial intro-differential equation.  In Cartesian coordinates the
$\nabla =  \frac{\partial}{\partial x} + \frac{\partial}{\partial y} +
\frac{\partial}{\partial z}$ and the direction $\hat{\vec n}$ is defined by two
angles $(\theta,\phi)$ at the position $\vec x$.
}

{
\subsection{The operator splitting method}

The mean intensity $J$ is obtained from the source function
$S$ by a formal solution of the RTE which is symbolically written
using the $\Lambda$-operator $\Lambda$ as
\begin{equation}
     J = \Lambda S.              \label{frmsol}
\end{equation}
The source function is given by $S=(1-\e)J + \e B$, where $\e$ 
denotes the thermal coupling parameter and $B$ is Planck's function.

The $\L$-iteration method, i.e.\ to solve Eq.~\ref{frmsol} by a fixed-point
iteration scheme of the form
\bea
   \Jnew = \L \Sold , \quad
   \Snew = (1-\e)\Jnew + \e B  ,\label{alisol}
\eea
fails in the case of large optical depths and small $\e$.

{Here, $\Jb = \int_0^\infty J_\lambda \phi_\lambda \, d\lambda$ is
  the mean intensity averaged over the line profile, $\phi_\lambda$; $\Sold$ is the current estimate for the source
function $S$; and $\Snew$ is the new, improved, estimate of $S$ for the next
iteration.} { The failure  of the $\Lambda$-iteration to converge} is caused
by the fact that the largest eigenvalue of the amplification matrix
is approximately \cite[]{mkh75}
$\l_{\rm max} \approx (1-\e)(1-T^{-1})$, where $T$ is the  optical 
thickness of the medium. For small $\e$ and large $T$, this is very close
to unity and, therefore, the convergence rate of $\L$-iteration is very 
poor. A physical description of this effect can be found in 
\cite{mih80}.

The idea of the operator splitting (OS) method is to reduce the 
eigenvalues of the amplification matrix in the iteration scheme
\cite[]{cannon73}  by 
introducing an approximate $\L$-operator (ALO) $\lstar$
and to split $\L$ according to
\begin{equation}
           \L = \lstar +(\L-\lstar) \label{alodef}
\end{equation}
and rewrite Eq.~\ref{alisol} as
\begin{equation}
     \Jnew = \lstar \Snew + (\L-\lstar)\Sold. 
\end{equation}
This relation can be written as \cite{hamann87}
\begin{equation}
    \left[1-\lstar(1-\e)\right]\Jnew = \Jfs - \lstar(1-\e)\Jold, \label{alo}
\end{equation}
where $\Jfs=\L\Sold$ { and $\Jold$ is the current
estimate of the mean intensity $J$}. Equation~\ref{alo} is solved to get the new values of 
$\Jb$ which is then used to compute the new 
source function for the next iteration cycle.
The
calculation and the structure of $\lstar$ should be simple in order to
make the construction of the linear system in Eq.~\ref{alo} fast. For
example, the choice $\lstar=\L$ is best in view of the
convergence rate (it is equivalent to a direct solution by matrix inversion)
but the explicit construction of $\L$ is more time
consuming than the construction of a simpler $\lstar$. 

The CPU time required for the solution of the RTE using the OS method depends
on several factors: (a) the time required for a formal solution and the
computation of $\Jfs$, (b) the time needed to construct $\lstar$, (c) the time
required for  the solution of Eq.~\ref{alo}, and (d) the number of iterations
required for convergence to the prescribed accuracy.
}

\section{Test cases }

The basic test setups for the Cartesian grid with periodic boundary 
conditions (PBCs) and the spherical coordinate system (SP) were 
chosen so that the tests can run on the smallest memory system 
(GPU) that we used here.  These setups are the basic building blocks
that are employed in more complex models (e.g., 3D NLTE calculations).

\subsection{Test case setup: periodic boundary conditions}

The test cases we have investigated follow the continuum tests used in
\cite{3drt_paper3} and \cite{3drt_paper8}. 
In detail, we used
a configuration that utilizes 
periodic boundary conditions (PBCs) in a plane
parallel slab. We used PBCs on the $x$ and $y$ axes, $z_{\rm max}$ is
the outside boundary, and $z_{\rm min}$ the inside boundary. The slab
has a finite optical depth in the $z$ axis. 
The basic model parameters are
\begin{enumerate}
\item the total thickness of the slab,  $z_{\rm max}- z_{\rm min} = 10^8\,$cm;
\item the minimum optical depth in the continuum, $\t_{\rm std}^{\rm min} =
10^{-4}$ and the maximum optical depth in the continuum, $\t_{\rm std}^{\rm
max} = 10^8$;
\item grey temperature structure for $\Teff=10^4$~K;
\item  boundary conditions with outer boundary condition $I_{\rm bc}^{-}
  \equiv 0$ and  
inner boundary condition LTE diffusion;
\item parameterized coherent and isotropic continuum scattering given by
\[
\chi_c = \epsilon_c \kappa_c + (1-\epsilon_c) \sigma_c
\]
with $\epsilon_c = 10^{-2}$. 
$\kappa_c$ and $\sigma_c$ are the 
continuum absorption and scattering coefficients.
\end{enumerate}
The Cartesian grid has $(n_x,n_y,n_z) = (65,65,257) = 1085825$ voxels
and we use $\Omega = (n_\theta,n_\phi) = (32,16) = 512$ solid angles
for the formal solution, equally spaced in $\mu = \cos(\theta)$ and
$\phi$. 

\subsection{Test case setup: spherical coordinate system}

The setup for the spherical coordinate system test (SP)
follows the general setup for the corresponding case
in Paper IV \citep[][]{3drt_paper4}. The configuration
used here is:
\begin{enumerate}
\item Inner radius $r_c=10^{10}\,$cm, outer radius $\rout = 2\alog{10}\,$cm.
\item Minimum optical depth in the continuum $\t_{\rm std}^{\rm min} =
10^{-8}$ and maximum optical depth in the continuum $\t_{\rm std}^{\rm
max} = 10^{1}$.
\item Grey  temperature structure with $\Teff=10^4$~K.
\item Outer boundary condition $I_{\rm bc}^{-} \equiv 0$ and diffusion
inner boundary condition.
\item Continuum extinction $\chi_c = C/r^6$, with the constant $C$
fixed by the radius and optical depth grids.
\item Parameterized coherent and isotropic continuum scattering by
defining
\[
\chi_c = \epsilon_c \kappa_c + (1-\epsilon_c) \sigma_c
\]
with $\epsilon_c = 10^{-2}$.
$\kappa_c$ and $\sigma_c$ are the
continuum absorption and scattering coefficients.
\end{enumerate}
The spherical grid (SP) has $(n_r,n_\theta,n_\phi) = (129,33,65) = 276705$ voxels
and we use $\Omega = (n_\theta,n_\phi) = (8,16) = 128$ solid angles
for the formal solution (equally spaced in $\mu = \cos(\theta)$ and
$\phi$. This is smaller than in typical \phxT\ applications, however,
this setup still fits into the smallest RAM of the test systems.
The timing profile of this setup is similar to large
scale \phxT\ runs, so that we use it as a model in this work.

\subsection{Test systems}

\subsubsection{CPU: Intel Xeon}

The base system for the comparison is a dual Intel Xeon CPU W-3223 
with  3.50GHz clock-speed at 8 cores per CPU and 2 hardware threads per core.
The system  is realized in a MacPro7,1 system with 96GB RAM. The OS is 
MacOS 10.15, the available compilers are GCC 10.2.0, Intel 19.1.3.301
and PGI 19.10-0 (the newer NVIDIA HPC compilers were not available for MacOS).

\subsubsection{GPU: NVIDIA V100}

The GPU test system is a 8 core Xeon Silver 4110 (Skylake) 2.1GHz host machine with a NVIDIA V100
GPU with 32GB RAM. The GPU has a clock speed of 1380\,MHz and a Memory
Clock Rate of 877\,MHz and 4096\,bits memory bus width. To target the GPU, we
use OpenACC as implemented in the NVIDIA Fortran compiler version 20.9-0.  The host
system is running Centos 7. For the host CPU the Intel compiler
version 19.1.3.304 is also
available.

\subsubsection{Many-core: KNL}

The many-core test system is a Intel Xeon Phi CPU 7250 (Knights
Landing; KNL)
based system with  1.30GHz clock speed, 64 cores and 4 hardware threads per
core, similar to the NERSC Cori nodes
(\url{https://www.nersc.gov/users/computational-systems/cori/}).  On this
system with use Intel compiler version 19.1.3.304 with OpenMP directives
turned on.

\subsubsection{Vector: NEC SX-Aurora TSUBASA}

As a vector processor test system we use a NEC SX-Aurora TSUBASA A101 (Aurora) machine with
2 vector engine (VE) cards. Each card has 48GB RAM, 8 vector cores at 1.4\,GHz
clock speed. The vector host (VH) is a Xeon Gold 6126 CPU (Skylake) at  2.60GHz
clock-speed with 96GB RAM. The VH manages the 2 VEs and handles their I/O
requests, runs the compiler etc.  The (MPI) application runs completely on the
VEs and is natively compiled using the NEC Fortran compiler, version 2.1.1.
The VE have fast vector pipelines and a scalar unit on each core. For the
test we use the NEC Fortran Compiler to create vectorized executables with the help of
compiler directives and NEC MPI for the MPI based parallelization. 

In addition to the VE, the VH can also be used for calculations. We thus ran
tests with the Intel Fortran Compiler (version 19.1.3.304) and the NVIDIA compiler
(version 20.9-0) with their respective MPI and OpenMP implementations.

\section{Method}

In the following discussion we use the notation of Papers I -- XI.

Considering the 3DRT module individually for any given wavelength, the biggest
time consumers are: setting up the geometric paths through the grid for all
solid angles (henceforth: the tracker), computing the formal solution
for all solid angles, computing the $\Lstar$ operator (only in the first
iteration) and the operator splitting solver (computing the new $J$'s by
solving the operator splitting linear system, see Paper I). In addition, MPI
load balancing may be measurable (e.g., if the workload for different solid angles is
significantly different so that some MPI processes finish faster than others,
this is the case in the PBC test setup). These steps are repeated
for all wavelength points that are considered in a calculation. 

For the solution of the large sparse linear system (its rank is equal to the
total number of voxels of the simulation) in the operator splitting step (`OS
iteration') we have implemented 3 algorithms: a Jacobi iteration, a
Gauss-Seidel iteration, and the BiCGStab method \citep{siam_templates}, parallelized with MPI, OpenMP
and OpenACC.  Each of them performs differently on the test systems and we show
in the graphs the fastest version for each of the test runs.

The standard algorithm implements the algorithms described in Paper III
(PBCs) and IV (SP) for multi-CPU systems with MPI parallelization: In the
formal solution (and $\Lstar$ computation) different solid angles are
parallelized and the data needs to be
collected from all participating processes only at the end of the
formal solution. Similarly, the operator splitting step is parallelized
with MPI.  The standard algorithm was 
designed to minimize the memory footprint, so that calculations can be performed on
smaller machines. One important observation is that the tracker and the formal solution are
memory, rather than compute bound: The different solid angles cause memory
access patterns that are aligned with optimal memory and
cache access patterns only in specific cases --- in most cases memory is accessed randomly through the
formal solution.  Thus, memory latency is the primary bottleneck.

Some results for the standard algorithm are shown in Fig. \ref{fig:standard}.
The performance on older Skylake (Silver 4110) and newer Cascade Lake (W-3223) CPUs is roughly the
same. All times are given for a complete node, i.e., all cores of the CPU(s)
of the node are used, a common pattern in High Performance Computing, (HPC). The standard algorithm is not
multi-threaded with OpenMP, thus the multiple CPU cores are used with MPI
parallelization. For comparison we also include the results on the Intel Xeon
Phi KNL many-core CPU and the NEC Aurora vector CPU. Whereas the standard
algorithm on the KNL CPU fares very well, its performance on the vector
processor is abysmal. This is due to the un-threaded and un-vectorizable form
of the standard algorithm. The theoretical performance of the KNL cannot be
realized in this setup since the standard algorithm does not produce
performance gains with OpenMP threading: It requires critical
sections and atomic constructs to give correct results and the memory
access pattern causes delays in the OpenMP threads, so that OpenMP directives
can even be counter productive. 

In the following, we will refer to the ``formal solution'' as the
combination of the formal solution and the the construction of the
nearest-neighbor $\Lstar$. Each of these steps, formal solution and
$\Lstar$ construction take about the same computation time. 
In all cases for the standard algorithm, the formal solution 
takes the largest time (red bars in 
Fig. \ref{fig:standard}), whereas the operator splitting solver (blue) 
timing varies from system to system. The time for the MPI `allreduce' (green)
is substantially larger in the PBC case, where the workload between different
MPI processes is significantly different, than in the SP model, where the
load balancing is much better. 

\subsection{Many-core algorithm}

As a first step towards better performance and constructing an optimized
algorithm for the formal solution on many-core CPUs we consider the tracker; in
particular, the geometry part. In a typical \phxT\ application, many (100,000
or more) wavelength points are considered for the same structure.  NLTE
modeling has to repeat this procedure numerous times in order to solve the
multi-level NLTE rate equations consistently with the radiation field.
Therefore, the geometry and thus, the set of tracks (directions in momentum
space) through the voxel grid remain the same for all wavelength points.  This
can be used to setup and fill a geometry cache once, at the beginning, which is
than accessed for all subsequent formal solutions, until  the  wavelength grid
has been traversed. This requires (a) a multi-pass algorithm which separates
the geometry tracking from the formal solution and (b) optionally storing the
geometry tracking data.  This requires substantial additional memory, which is,
however, fully distributed over different MPI processes with no communication
and sharing required. Thus, when using more MPI processes, each process needs
to progressively store less geometry data up to the theoretical limit of one
solid angle per MPI process. This situation is actually very realistic in
\phxT\ simulations as the domain decomposition requires anywhere from 64 to
1024 MPI processes (typically 4-64 nodes) and in such realistic setups, the
geometry cache requires only a small fractional memory increase per MPI
process.

The multi-pass algorithm has the additional advantage that it allows for more
efficient OpenMP parallelization and Single Instruction Multiple Data
(SIMD) style vectorization. Once 
the geometric paths through the voxel grid are known (either by computing them
in a first phase or by recalling them from the geometry cache) for each solid
angle, the formal solution can be computed by loops over the voxel grid rather
than stepping along each characteristic.  To keep the loops short and simple
enough for OpenMP based parallelization, it turned out to be advantageous to
split the calculation into separate phases. Each phase can then use 
different OpenMP parallelization and SIMD statement vectorization (for Intel
CPUs, using the Intel compilers). This increases memory locality, resulting
in better cache usage, but does add temporary arrays that are needed to store
intermediate results. 

For a multi-core CPU the results are shown in  Fig. \ref{fig:Xeon}, where we
use the  Xeon W-3223 as an example.  Here, the multi-pass algorithm without the
geometry cache (labeled `MultiPass' in the figures) performs substantially better
than the standard algorithm  (about 1.9 times) in the PBC case but
substantially \emph{worse} (by a factor of 2.4) in the SP case.  This is caused
by the more complex data structures required by the MultiPass algorithm where
the geometry tracking phase cannot overlap with the computation phase.
However, with the geometry cache active, the results change substantially. Note
that in order to simulate the typical results obtained within a larger \phxT\
model the times do \emph{not} include the construction of the geometry
cache (this is done once before the calculations start). This `MultiPassCache'
version produces the best performance for both test cases, about 3 times
speedup for the SP case and 2.6 times for the PBC test compared to the standard
algorithm.  These are substantial speedups on classic multi-core CPUs.  Listing
\ref{lst:openmp} gives pseudo-code for this method.

The corresponding results on many-core CPUs are similar. In Figs. \ref{fig:KNL}
and \ref{fig:KNL2} we show the results obtained on the Xeon Phi 7210 (64 cores
with 4 hardware threads each at 1.3GHz).  Here, the standard algorithm is
substantially slower than the `MultiPassCache' setup. The speedups for the
MultiPass+Cache version are 3.4 (PBCs) and 2.6 (SP) compared to the standard
algorithm. Enabling the geometry cache speeds up the overall calculation by a
factor of 2 or more. For a full \phxT\ simulation run this produces 
massively reduced wallclock times
compared to the standard and MultiPass algorithms.  In the case of
KNL, the MultiPassCache algorithm allows for much better OpenMP
utilization, including the OpenMP SIMD statement that is used to vectorize
(i.e., use AVX-512 instructions) on the KNL CPU with the Intel compiler. These
vectorizations by themselves already produce significant speed-ups compared to
the standard algorithm.  This result is highlighted by the observation (cf.
Fig.~ \ref{fig:KNL2}) that for the SP test on KNL, a 16\,MPI processes with
16\,OpenMP threads each (16@16)  setup, is faster than the more MPI biased 64@4
setup. Even the very openMP focused 4@64 setup is only 10\% slower than the
16@16 setup.  In the PBC test case, the 4@64 setup is the fastest; however, the
16@16setup is only about 12\% slower.  This is likely at least, in part, due to
MPI load balancing, which is a bigger problem in the PBC test (where the
different characteristics have very different lengths in terms of voxel counts
than the far more balanced SP test case).  These effects show that the KNL is
quite sensitive to details of the setup and  for each simulation the optimal
setup may need to be determined by testing in advance.  Experiments using less
than the 4 hardware threads per core on KNL (as suggested in some
performance documents) reduced performance significantly.  It appears that
using more threads may be able to hide memory latencies better; however, using
more than 4 threads per core also reduced performance.

\subsection{Vector processor algorithm}

The Aurora is a vector processor with significantly different performance
characteristics than Intel Xeons or many-core CPUs.  While its vector
performance is very high, its scalar performance is quite low, that is,
vectorization is of utmost importance on this system.  In  Fig. \ref{fig:NEC}
we show the performance results for the standard and MultiPass algorithms,
which do \emph{not} use the vector processing capabilities of the Aurora and
the execution times are very long. 

Therefore, we have developed a variant of the MultiPass+Cache algorithm that
can be vectorized (with directives) by the NEC Fortran compiler.  For this, we
swapped loop orders (adding intermediate helper arrays where necessary) and
split/merged loops so that the NEC compiler vectorized them. The design goal
was to vectorize the longest possible loop, even if it requires additional
arrays to store intermediate data. The resulting code uses slightly more RAM
than the MultiPass+Cache version.  In addition, we also developed a vectorized
version of the operator splitting solver in order to further increase overall
performance. The changes required to the standard operator splitting solvers
are small, a few loop rearrangements and vectorization directives.

The performance of the vector algorithm is shown in  Fig. \ref{fig:NEC}.
Compared to the standard algorithm, the vector version on the Aurora is about
44 times faster for the PBC case and 65 times faster for the SP test case. In
addition, the vector version significantly reduces MPI load balance issues. On
the multi-core CPU the vector algorithm is 1.5 times faster than the standard
algorithm in the PBC case, but only about 17\% faster compared to the standard
algorithm for the SP test case.  On the many-core Xeon Phi, the vector
algorithm is 3-4 times \emph{slower} than the MultiPass+Cache version (not
shown). Note that the vector algorithm does not use OpenMP threading, so that,
in particular, on KNL, only one thread per core is used.  In addition, the
Intel compiler does \emph{not} vectorize the loops in the same way that the NEC
compiler does and, therefore, the vector algorithm does \emph{not} use the Xeon
vector instructions (e.g., AVX-512 on the KNL) efficiently.

\subsection{GPU (OpenACC) algorithm}

Over the last decade, using GPUs to accelerate numerical calculations (compared
to classical CPUs) has become important. To enable such usage, proprietary
methods \cite[e.g., CUDA,][]{cuda} as well as open standards have been
developed. The latter are realized both as programming standards, e.g., OpenCL
\citep[][]{OpenCL}, and as (directive based) APIs, e.g., OpenACC
\citep[][]{OpenACC}.  For long term portability and vendor independence, we
consider it imperative to utilize  an open standard based approach (codes are
often used  for decades on very different hardware generations).  We have
published results for OpenCL before \citep[][]{3drt_paper8} and therefore we
concentrate here  on an OpenACC based approach.  New versions of OpenMP also
support GPU based offloading; however, the support for available hardware
provided by existing compilers that we have access to is presently very
limited. Currently, this is actually also true for OpenACC, where the
only compiler support available is the NVIDIA compiler \citep[][]{NVIDIAHPC} on
NVIDIA hardware. OpenACC support in GCC \citep[][]{GCC} was, at the time of
this writing, not in an advanced enough state to be usable.

For the OpenACC version we had to make significant changes to the code.  The
main problem is that OpenACC does not work with the array-of-structures method
that the original code (standard, cache, and vector versions) uses.  Therefore,
we had to redesign the code to also (alternatively) use a structure-of-arrays
method (labeled `FlatCache' in the figures) to store the geometry tracking
cache and the data used for the operator splitting (e.g., the $\Lstar$ array).
On the NEC and the Xeons, the structure-of-arrays version is marginally
faster (about 2\%) than the array-of-structures method.
The individual arrays are then easily transferred onto the GPU with OpenACC
update directives.  If the GPU device has enough memory, the geometry cache can
be stored once on the GPU and then be used directly, without the need to
transfer the tracking cache for each solid angle. In the test cases, the caches
and additional 3DRT arrays are 25-30GB total, so that they can be stored on the
GPU (LargeGPU mode). For comparison, we have also run tests in Small GPU mode
where the tracking caches are kept on the host (CPU) and transferred onto the
GPU for each solid angle. In this mode, about 1-2GB are used on the GPU so that
smaller devices can also be used or multiple processes can be run on one GPU.
Note that in MPI mode with several GPUs only a fraction of the tracking cache
needs to be stored, thus, proportionally reducing the memory use per GPU. 

The OpenACC version of the formal solution is adapted from the 
vector version by adding OpenACC directives. In a few loops, the 
order was exchanged to better exploit the GPU architecture. 

Listing \ref{lst:openacc} gives the pseudo-code for this method. The
compiler directives are relatively simple and converting from 
OpenMP to OpenACC is relatively straightforward. Likewise, given a
working OpenACC version it is relatively straightforward to port to
OpenMP based device directives, based on small test cases that we have
been able to evaluate.

The results of the OpenACC test calculations are shown in Fig. \ref{fig:PGI},
where we compare the timings obtained with the NVIDIA/OpenACC compiler on the
GPU. NVIDIA OpenACC support is also available for many-core CPUs, thus we are able
to include results for the KNL used as an OpenACC device with shared memory. 
The test GPU is a NVIDIA V100 with 32\,GB, which can be run in both
SmallGPU and LargeGPU mode. The LargeGPU mode is 4 to 9 times faster
than the SmallGPU mode, clearly showing that data transfer is a significant 
bottleneck for the GPU. The KNL OpenACC code is, in comparison, much slower
and not competitive with the MPI+OpenACC versions (not shown).

\subsection{Xeon compiler comparison}

The timing results for the same jobs on the Xeon W-3223 CPU (this is a
MacPro7,1 running MacOS, and when the PGI compiler is used, the NVIDIA HPC
SDK is not 
available for MacOS) different compilers are shown in Fig. \ref{fig:compiler}.
For both test setups we used the  MultiPass+Cache algorithm with MPI+OpenMP (8
processes with 2 threads each, i.e., hyperthreaded mode) on the system, the
fastest results for each compiler are shown in Fig. \ref{fig:compiler}.  The
compilers are Intel Version 19.1.3.301, GNU Fortran (GCC) 10.2.0, and PGI
pgfortran 19.10-0. In both model setups, the Intel compiler results in the
smallest run time, followed by GCC which is about 12\% slower in the SP test
and 42\% slower in the PBC test compared to the Intel compiler results. The PGI
compiler trails in both tests significantly, a test with the newer NVIDIA HPC
compiler on the KNL and the Xeon Silver 4110 (host CPU of the V100 GPU) shows
similar trends.

\section{Summary and conclusions}

In Fig.~\ref{fig:total} we show the results for the fastest 7 runs over all
systems, algorithms, and setups combined. For the PBC test case, the OpenACC
LargeGPU version on the V100 GPU is by far the fastest, followed by
the NEC TSUBASA Aurora vector
processor which is slower by a factor of about 3.2. In the spherical test case
the OpenACC LargeGPU setup is also the fastest, followed by the NEC
TSUBASA Aurora which is a
factor of about 2.4 slower. The 5 year old KNL is surprisingly quick, it beats
the V100 in SmallGPU mode for the PBC tests and is faster than the newer
multi-core Xeons. In all cases, the geometry caching is very effective,
producing significant speedups, particularly, on more recent CPUs (for
example, a Skylake Xeon Gold 6126 is 1.89 faster with the cache enabled than
without it). The OpenACC SmallGPU version is much slower than the LargeGPU
setup, showing the cost of data transfer. This is of great practical importance
as large scale 3DRT runs require more RAM than current GPUs have available, in
which case either SmallGPU mode or multiple GPUs must be used. In SmallGPU
mode, the overall utilization of the GPU is smaller (by about 50\%) so that
multiple parallel processes on a single GPU may be used, initial experiments have
provided promising results for very large test cases and 2 independent processes
on a single V100 GPU (using NVIDIA MPS).

The main problems with OpenACC are that the compiler support needed for
efficient and portable (to different GPU vendors) code is not yet available and
that complex code and data structures may need to be adapted (simplified) for
efficient OpenACC data transfer.  Overall, OpenACC code is simple to
generate using the vector algorithm as  a starting point. In the future it may
be better to switch to the OpenMP offload/target paradigm which appears to be
available for more hardware options and is supported by more
compilers, in particular the LLVM framework\footnote{\url{llvm.org}}.

The NEC Aurora vector processor is very fast  and easy to code
for, however, its scalar performance is very low. In practical applications it
may be best to combine the vector engine with the host CPU whenever possible,
where the 3DRT executes on the vector processor and the host CPU is used for
I/O and scalar processing. Several methods allow for this option, including
running MPI processes on the vector processor and the host CPU at the same time
and offloading to and from the vector processor.

The performance of the  Xeon Phi KNL is quite sensitive to the exact setup (MPI
and OpenMP) used for each problem, thus, in typical  production use it is
important to determine the optimal configuration beforehand (or to implement an
automatic scheme to optimize the configuration at runtime). On all Intel CPUs,
the vector algorithm is less efficient than the MultiPass+Cache version
(recall that the vector version is a rearranged MultiPass version and
includes the geometry cache). This could be due to the Intel 
compiler not using the vector instructions (which is forced by OpenMP
SIMD directives in the MultiPass algorithm) automatically or by the vector
instructions stalling to data gather or scatter (which may be hidden
by the many threads used on the Xeon Phi).

Our results enable much larger model calculations than were previously
feasible. On standard CPU hardware the algorithms developed for the KNL give at
least a factor of 3--4 speedup over standard Xeons, if NEC vector or GPU
hardware is available speedup factors of 7-21 are possible.  As the 3DRT takes
75\% of the total simulation time, this speedup can reduce the overall runtime
by up to 75\%, a massive savings in computer time.  Simulations that took a year
before, will now take a mere quarter year and/or can be run on much smaller
systems. Supercomputers with the required hardware are already available, for
example, Summit, or will become available soon, e.g., NERSC's next generation
Perlmutter system. 

Future work will include several instances of solar-type stars with
chromospheres so that we can define a 3D model of the quiet sun, pre-CV stars
in 3D, core collapse and thermonuclear supernovae interacting with their
environment, and models of neutron star mergers.

\subsection*{Acknowledgments}
PHH gratefully acknowledges support by the DFG under grants HA 3457/20-1 and HA
3457/23-1.  Some of the  calculations presented here were performed at the RRZ
of the Universit\"at Hamburg, at the H\"ochstleistungs Rechenzentrum Nord
(HLRN), and at the National Energy Research Supercomputer Center (NERSC), which
is supported by the Office of Science of the U.S.  Department of Energy under
Contract No. DE-AC03-76SF00098.  We thank all these institutions for a generous
allocation of computer time.  PHH gratefully acknowledges the support of NVIDIA
Corporation with the donation of a Quadro P6000 GPU used in this research.
E.B. acknowledges support from NASA Grant NNX17AG24G and an AUFF sabbatical
grant. We thank Dr. Rudolf Fischer from NEC Deutschland GmbH for very helpful
advise on vectorizing for the NEC TSUBASA Aurora system.

\bibliography{yeti,radtran,rte_paper2,stars,OpenCL,paper9,compsci}

\begin{thebibliography}{10}
\expandafter\ifx\csname url\endcsname\relax
  \def\url#1{\texttt{#1}}\fi
\expandafter\ifx\csname urlprefix\endcsname\relax\def\urlprefix{URL }\fi
\expandafter\ifx\csname href\endcsname\relax
  \def\href#1#2{#2} \def\path#1{#1}\fi

\bibitem{3drt_paper1}
P.~H. {Hauschildt}, E.~{Baron}, {A 3D radiative transfer framework. I.
  Non-local operator splitting and continuum scattering problems}, \aap 451
  (2006) 273--284.
\newblock \href {http://arxiv.org/abs/astro-ph/0601183}
  {\path{arXiv:astro-ph/0601183}}, \href
  {https://doi.org/10.1051/0004-6361:20053846}
  {\path{doi:10.1051/0004-6361:20053846}}.

\bibitem{3drt_paper2}
E.~{Baron}, P.~H. {Hauschildt}, {A 3D radiative transfer framework. II. Line
  transfer problems}, \aap 468~(1) (2007) 255--261.
\newblock \href {https://doi.org/10.1051/0004-6361:20066755}
  {\path{doi:10.1051/0004-6361:20066755}}.

\bibitem{3drt_paper3}
P.~H. {Hauschildt}, E.~{Baron}, {A 3D radiative transfer framework. III.
  Periodic boundary conditions}, \aap 490 (2008) 873--877.
\newblock \href {http://arxiv.org/abs/0808.0601} {\path{arXiv:0808.0601}},
  \href {https://doi.org/10.1051/0004-6361:200810239}
  {\path{doi:10.1051/0004-6361:200810239}}.

\bibitem{3drt_paper4}
P.~H. {Hauschildt}, E.~{Baron}, {A 3D radiative transfer framework. IV.
  Spherical and cylindrical coordinate systems}, \aap 498 (2009) 981--985.
\newblock \href {http://arxiv.org/abs/0903.1949} {\path{arXiv:0903.1949}},
  \href {https://doi.org/10.1051/0004-6361/200911661}
  {\path{doi:10.1051/0004-6361/200911661}}.

\bibitem{3drt_paper5}
E.~{Baron}, P.~H. {Hauschildt}, B.~{Chen}, {A 3D radiative transfer framework.
  V. Homologous flows}, \aap 498 (2009) 987--992.
\newblock \href {http://arxiv.org/abs/0903.2486} {\path{arXiv:0903.2486}},
  \href {https://doi.org/10.1051/0004-6361/200911681}
  {\path{doi:10.1051/0004-6361/200911681}}.

\bibitem{3drt_paper6}
P.~H. {Hauschildt}, E.~{Baron}, {A 3D radiative transfer framework. VI.
  PHOENIX/3D example applications}, \aap 509 (2010) A36+.
\newblock \href {http://arxiv.org/abs/0911.3285} {\path{arXiv:0911.3285}},
  \href {https://doi.org/10.1051/0004-6361/200913064}
  {\path{doi:10.1051/0004-6361/200913064}}.

\bibitem{3drt_paper7}
A.~M. {Seelmann}, P.~H. {Hauschildt}, E.~{Baron}, {A 3D radiative transfer
  framework . VII. Arbitrary velocity fields in the Eulerian frame}, \aap 522
  (2010) A102+.
\newblock \href {http://arxiv.org/abs/1007.3419} {\path{arXiv:1007.3419}},
  \href {https://doi.org/10.1051/0004-6361/201014278}
  {\path{doi:10.1051/0004-6361/201014278}}.

\bibitem{3drt_paper8}
P.~H. {Hauschildt}, E.~{Baron}, {A 3D radiative transfer framework. VIII.
  OpenCL implementation}, \aap 533 (2011) A127+.
\newblock \href {https://doi.org/10.1051/0004-6361/201117051}
  {\path{doi:10.1051/0004-6361/201117051}}.

\bibitem{3drt_paper9}
D.~{Jack}, P.~H. {Hauschildt}, E.~{Baron}, {A 3D radiative transfer framework.
  IX. Time dependence}, \aap 546 (2012) A39.
\newblock \href {http://arxiv.org/abs/1209.5788} {\path{arXiv:1209.5788}},
  \href {https://doi.org/10.1051/0004-6361/201118152}
  {\path{doi:10.1051/0004-6361/201118152}}.

\bibitem{3drt_paper10}
E.~{Baron}, P.~H. {Hauschildt}, B.~{Chen}, S.~{Knop}, {A 3D radiative transfer
  framework. X. Arbitrary velocity fields in the comoving frame}, \aap 548
  (2012) A67.
\newblock \href {http://arxiv.org/abs/1210.6679} {\path{arXiv:1210.6679}},
  \href {https://doi.org/10.1051/0004-6361/201219343}
  {\path{doi:10.1051/0004-6361/201219343}}.

\bibitem{3drt_paper11}
P.~H. {Hauschildt}, E.~{Baron}, {A 3D radiative transfer framework. XI.
  Multi-level NLTE}, \aap 566 (2014) A89.
\newblock \href {http://arxiv.org/abs/1404.4376} {\path{arXiv:1404.4376}},
  \href {https://doi.org/10.1051/0004-6361/201423574}
  {\path{doi:10.1051/0004-6361/201423574}}.

\bibitem{F2008}
"{Fortran Working Group}", Fortran 2008 Standard,
  \url{https://wg5-fortran.org/f2008.html} (2019).

\bibitem{MPI}
"{MPI Working Group}", The MPI Application Programming Interface,
  \url{https://www.mpi-forum.org/docs/} (2019).

\bibitem{OpenMP}
"{OpenMP Working Group}", The OpenMP Application Programming Interface,
  \url{https://www.openmp.org/specifications/} (2019).

\bibitem{OpenACC}
"{OpenACC Working Group}", The OpenACC Application Programming Interface,
  \url{https://www.openacc.org/specification} (2019).

\bibitem{aliperf}
P.~H. {Hauschildt}, H.~{St{\"o}rzer}, E.~{Baron}, {Convergence properties of
  the accelerated {\ensuremath{\Lambda}}-iteration method for the solution of
  radaitive transfer problems.}, \jqsrt 51~(6) (1994) 875--891.
\newblock \href {https://doi.org/10.1016/0022-4073(94)90018-3}
  {\path{doi:10.1016/0022-4073(94)90018-3}}.

\bibitem{mkh75}
D.~{Mihalas}, P.~B. {Kunasz}, D.~G. {Hummer}, {Solution of the comoving-frame
  equation of transfer in spherically symmetric flows. I. Computational method
  for equivalent-two-level-atom source functions.}, \apj 202 (1975) 465--489.
\newblock \href {https://doi.org/10.1086/153996} {\path{doi:10.1086/153996}}.

\bibitem{mih80}
D.~{Mihalas}, {Solution of the comoving-frame equation of transfer in
  spherically symmetric flows. VI - Relativistic flows}, \apj 237 (1980)
  574--589.
\newblock \href {https://doi.org/10.1086/157902} {\path{doi:10.1086/157902}}.

\bibitem{cannon73}
C.~J. {Cannon}, {Angular quadrature perturbations in radiative transfer
  theory.}, \jqsrt 13~(7) (1973) 627--633.
\newblock \href {https://doi.org/10.1016/0022-4073(73)90021-6}
  {\path{doi:10.1016/0022-4073(73)90021-6}}.

\bibitem{hamann87}
W.-R. Hamann, Line formation in expanding atmospheres: Multi-level calculations
  using approximate lambda operators, in: W.~Kalkofen (Ed.), Numerical
  Radiative Transfer, Cambridge University Press, 1987, p.~35.

\bibitem{siam_templates}
{Richard Barrett}, {Michael Berry}, {Tony F. Chan}, {James Demmel}, {June
  Donato}, {Jack Dongarra}, {Victor Eijkhout}, {Roldan Pozo}, {Charles Romine},
  {Henk van der Vorst}, Templates for the Solution of Linear Systems: Building
  Blocks for Iterative Methods, SIAM, Philadelphia, 1994, iSBN:
  978-0-89871-328-2.
\newblock \href {https://doi.org/https://doi.org/10.1137/1.9781611971538}
  {\path{doi:https://doi.org/10.1137/1.9781611971538}}.

\bibitem{cuda}
NVIDIA, Nvidia cuda: Compute unified device architecture, Tech. rep., NVIDIA
  Corp (2007).

\bibitem{OpenCL}
A.~Munshi, The opencl specification, Tech. rep., Khronos Group,
  http://www.khronos.org/registry/cl/specs/opencl-1.0.29.pdf (2009).

\bibitem{NVIDIAHPC}
NVIDIA, NVIDIA HPC SDK, \url{https://developer.nvidia.com/hpc-sdk} (2020).

\bibitem{GCC}
"{GNU Project}", GNU Compiler Suite Manuals,
  \url{https://gcc.gnu.org/onlinedocs/} (2019).

\end{thebibliography}

\clearpage

\begin{lstlisting}[float,basicstyle=\footnotesize,language=Fortran,caption={Pseudocode
for KNL OpenMP Implementation},label=lst:openmp]
!$omp parallel do default(none) collapse(2)
!$omp& private(various_variables)
!$omp& firstprivate(various_variables)
!$omp& shared(various_variables)
!
      do iz=-nz,nz  ! phase 3 loop:
       do iy=-ny,ny
!
!$omp simd
        do ix=-nx,nx
!--
!-- 0. step: the number of slots used is saved
!--
         lngth = chars_pv(ix,iy,iz)
!--
!-- 1. step: gather the previous and next points
!--          and the path lngths into vectors
!--
         i_loop: do i=1,lngth
!--
!-- 2. step: gather the previous and next Source and Opacities
!-- 3. step: compute dtau, dtau1
!-- 4. step: compute alpha, beta, gamma
!-- 4a. Fix up for BCs and vacuum
!-- 4b. store data for next phase if do_lstar is set
!-- 5. step: compute Delta I and exp(dtau)
!--
         enddo i_loop
        enddo
       enddo
      enddo ! phase 3 loop
!$omp end parallel do
\end{lstlisting}

\begin{lstlisting}[float,firstline=1,lastline=42,basicstyle=\footnotesize,language=Fortran,caption={Pseudocode
for OpenACC
Implementation},label=lst:openacc]
!--
!-- transfer data to the accelerator
!--
!$acc enter data
!$acc& create(various_variables)
!$acc update device(various_variables)
!
!$acc parallel loop collapse(3)
!$acc& present(various_variables)
!
      do iz=-nz,nz  ! phase 3 loop:
       do iy=-ny,ny
        do ix=-nx,nx
!--
!-- 0. step: the number of slots used is saved
!--
!--
!-- 1. step: gather the previous and next points
!--          and the path lngths into vectors
!--
!$acc loop seq
         i_loop: do i=1,lngth
!--
!-- 2. step: gather the previous and next Source and Opacities
!-- 3. step: compute dtau, dtau1
!-- 4. step: compute alpha, beta, gamma
!-- 4a. Fix up for BCs and vacuum
!-- 4b. store data for next phase if do_lstar is set
!-- 5. step: compute Delta I and exp(dtau)
!--
         enddo i_loop
        enddo
       enddo
      enddo ! phase 3 loop
!$acc exit data finalize
!$acc& delete(various_variables)
\end{lstlisting}

\def\expl{The horizontal bars give the overall  execution
walltimes (summed over all iterations) of the different phases of the
3D radiative transfer solver as 
indicated by the colors. `Formal Solution' designates the phase where new mean
intensities are computed from the current estimate of the source function
(including the construction of the $\Lstar$ operator in the first iteration),
`allreduce' is the phase where the contribution of all MPI processes are
collected and summed up via MPI functions and `OS iteration' is the time spent
computing the new estimate of the source function (including the solution of
the large sparse linear system). The labels on the right hand of the bars give
the overall execution time. The labels on the left hand side specify the system
or CPU, the algorithm, the solver of the large linear system in the OS step (Jacobi,
Gauss-Seidel oder BiCGStab), the compiler and the parallelization setup
(where the notation $x$@$y$ indices $x$ MPI processes with $y$ OpenMP
thread each).
}

\begin{figure*}
  \centering
\includegraphics[scale=0.25]{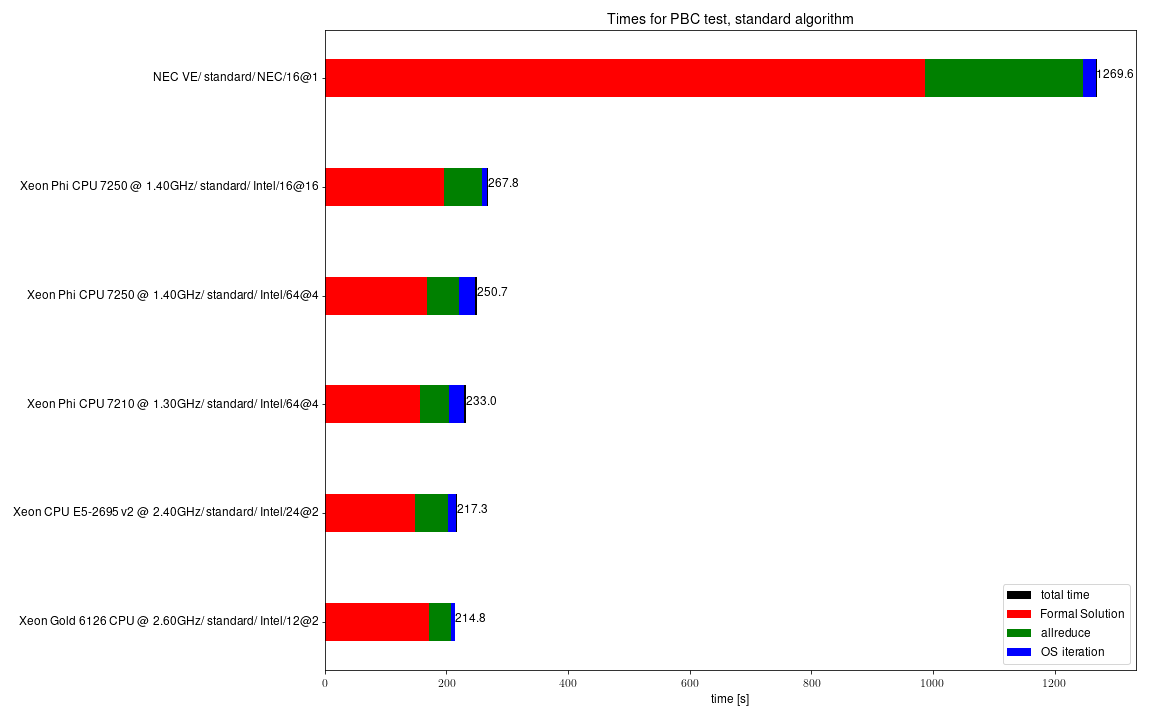}\\
\includegraphics[scale=0.25]{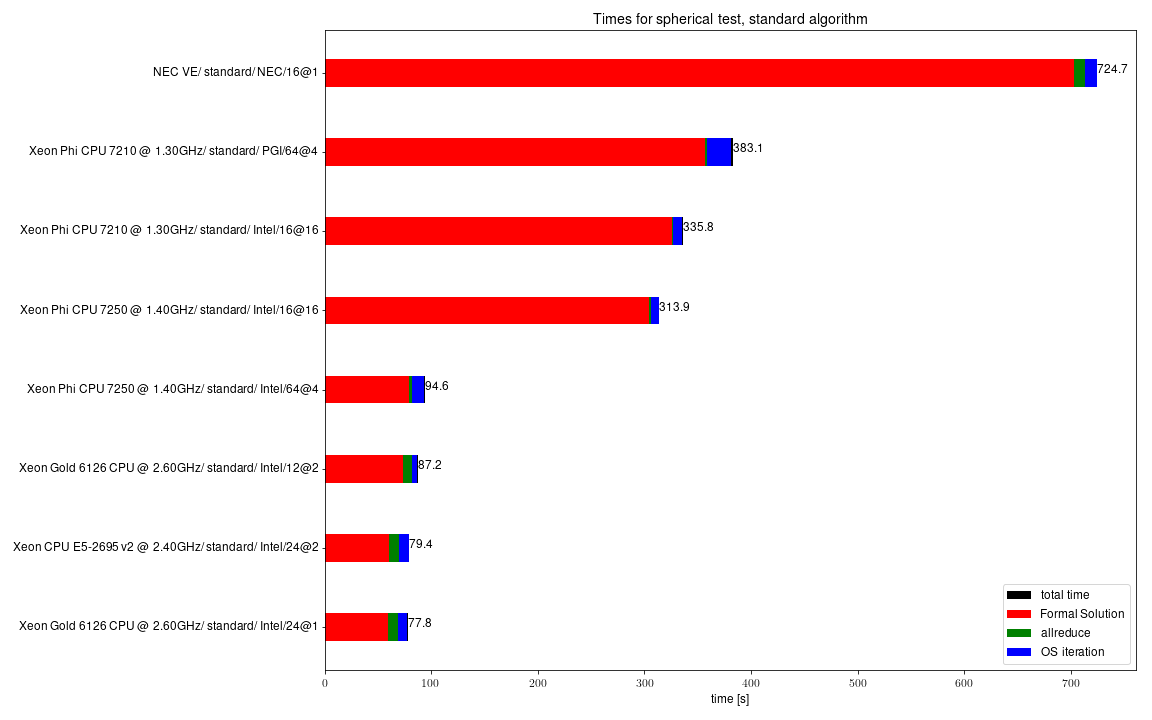}
\caption{\label{fig:standard}
Timing of the standard algorithm. \expl
}
\end{figure*}

\begin{figure*}
  \includegraphics[scale=0.25]{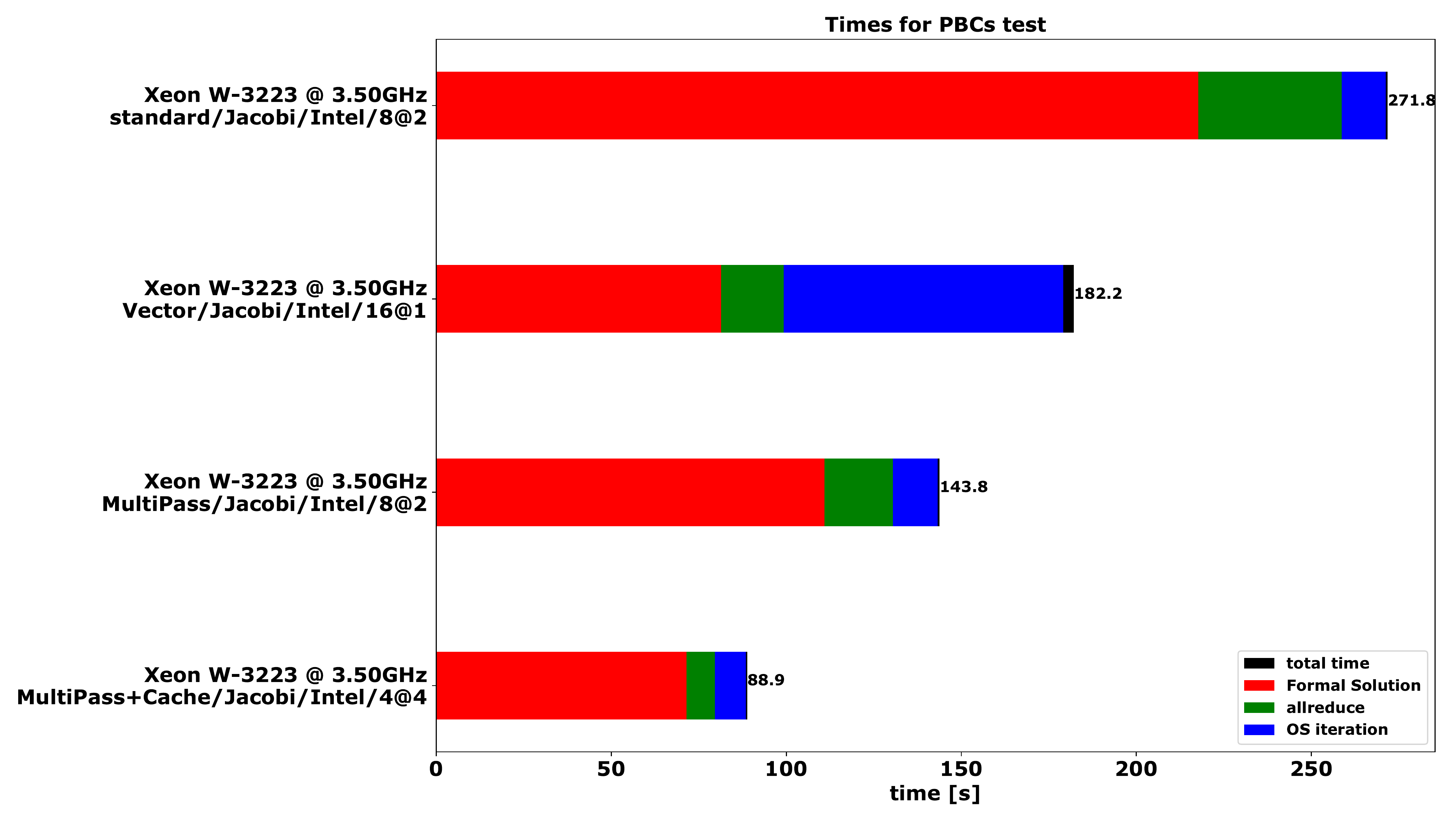}\\
\includegraphics[scale=0.25]{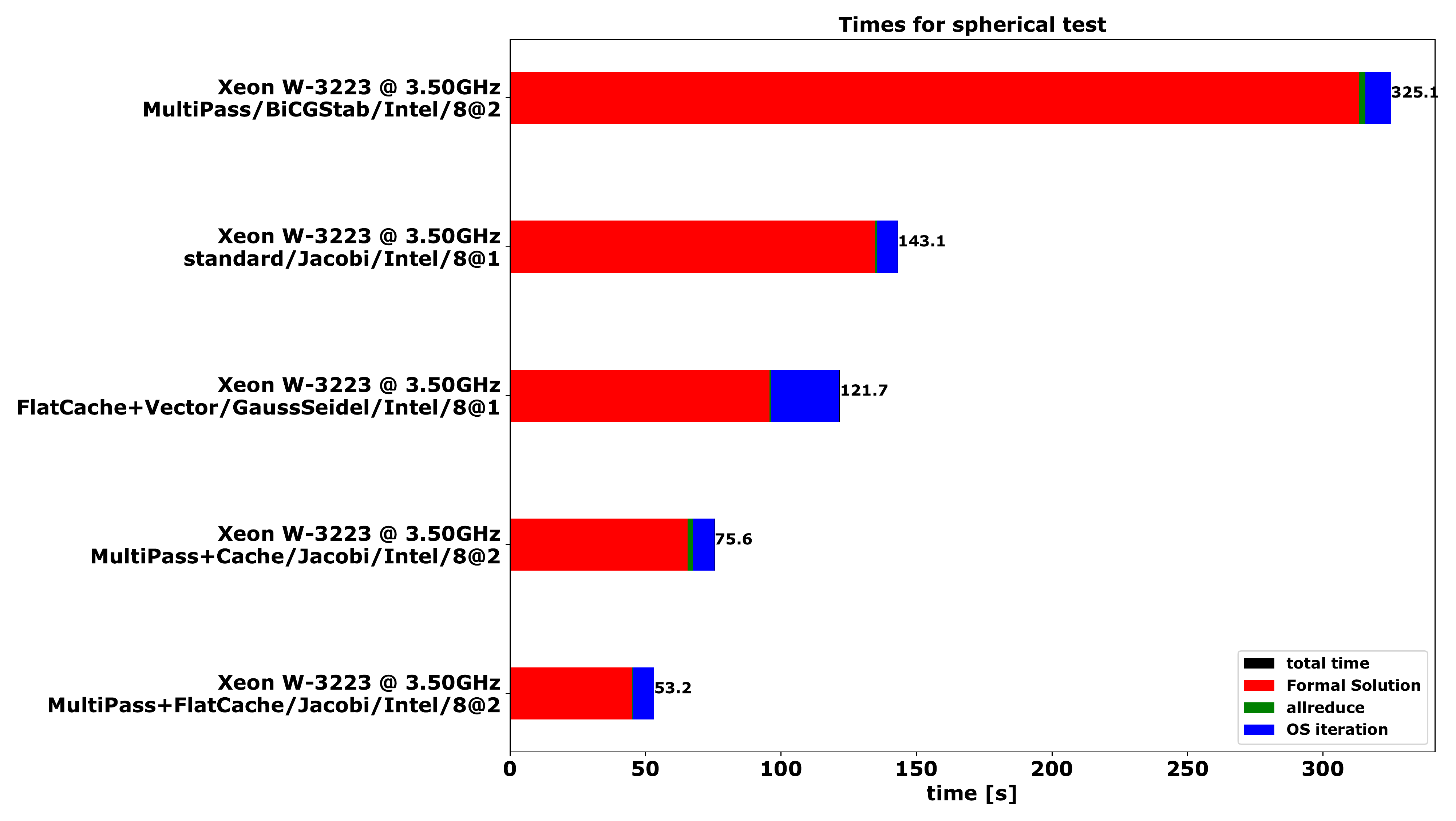}
\caption{\label{fig:Xeon}
Timing of different  algorithms on a modern many-core CPU (Xeon W-3223).
\expl
}
\end{figure*}

\begin{figure*}
\includegraphics[scale=0.20]{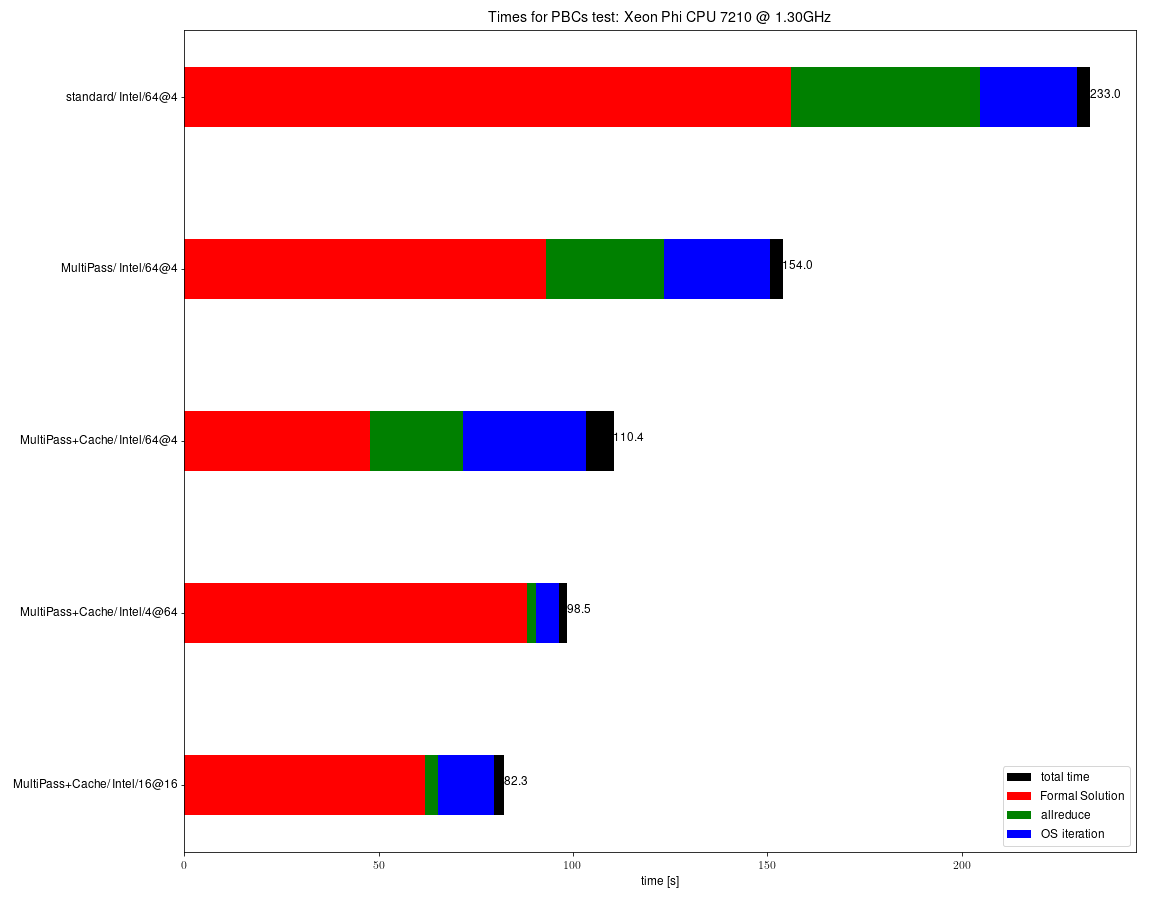}\\
\includegraphics[scale=0.20]{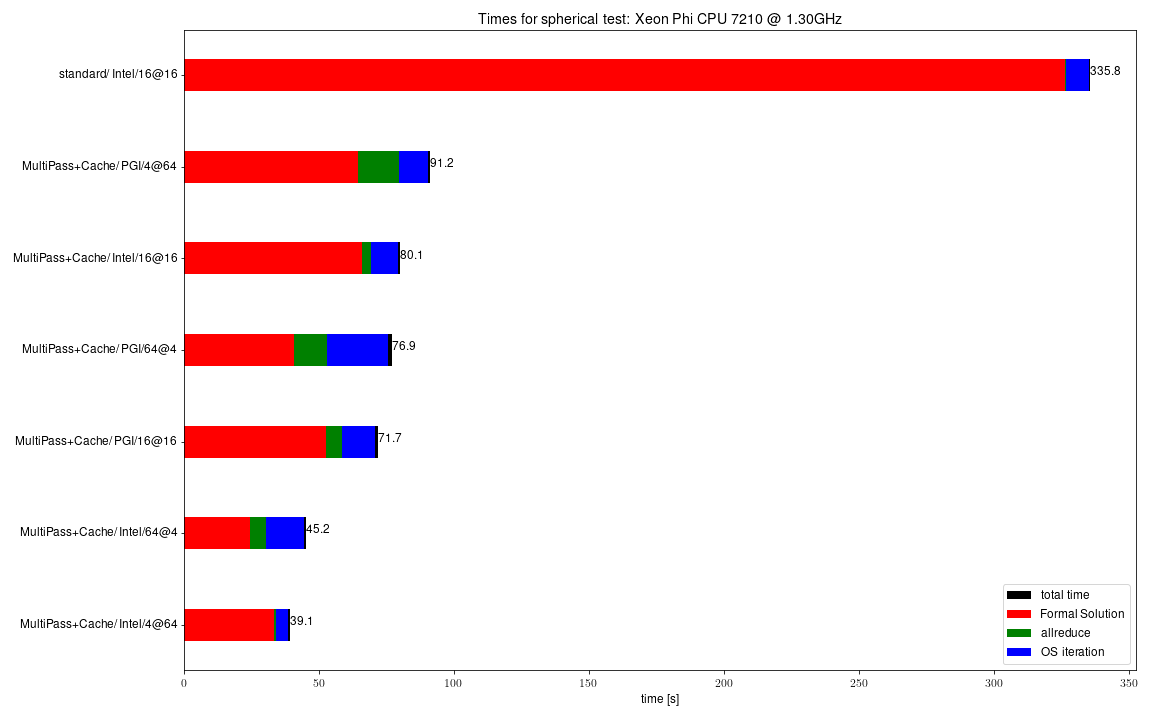}
\caption{\label{fig:KNL}
Timing of different  algorithms on a many-core CPU (Xeon Phi 7210).
\expl
}
\end{figure*}

\begin{figure*}
\includegraphics[scale=0.25]{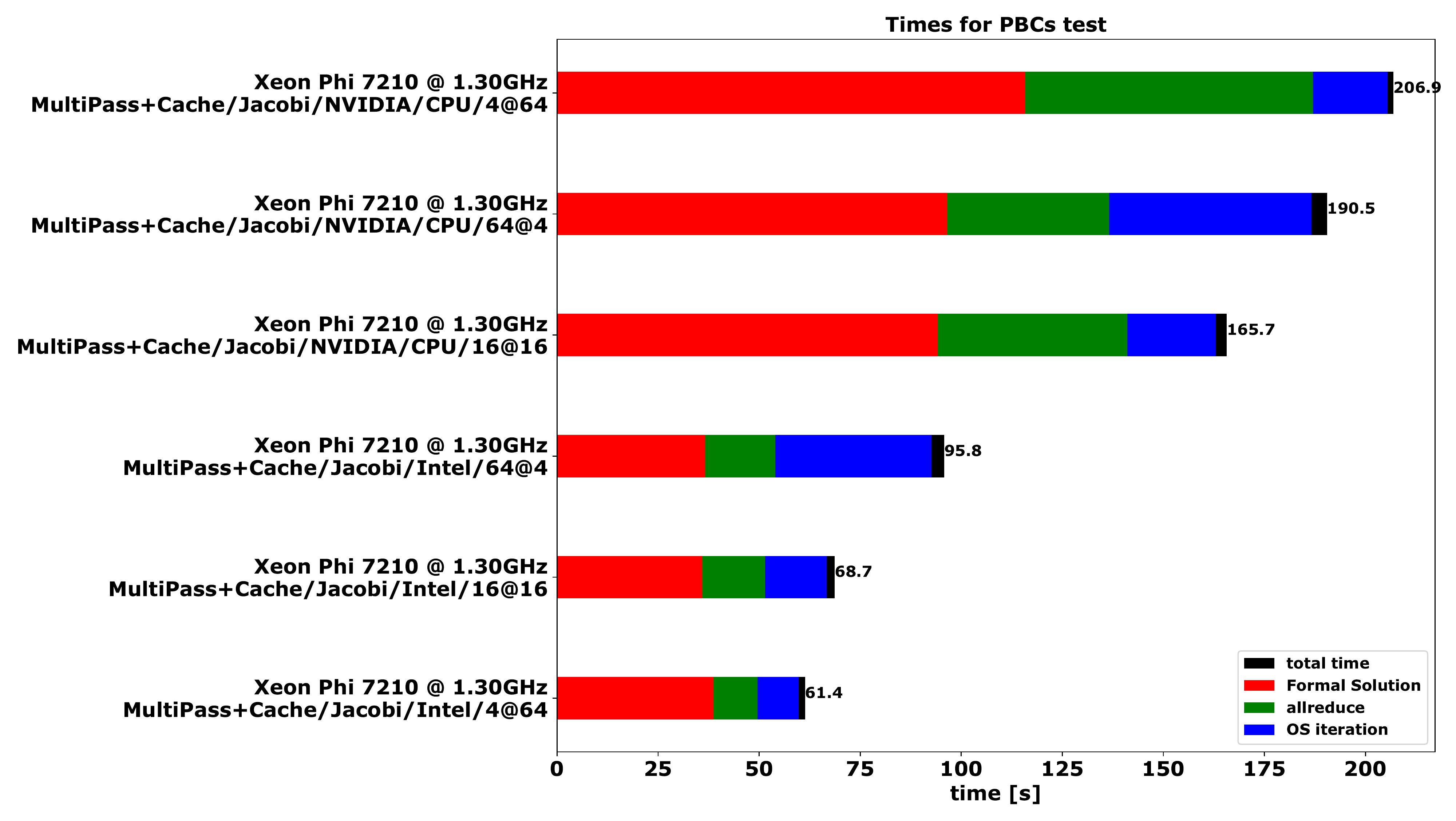}\\
\includegraphics[scale=0.25]{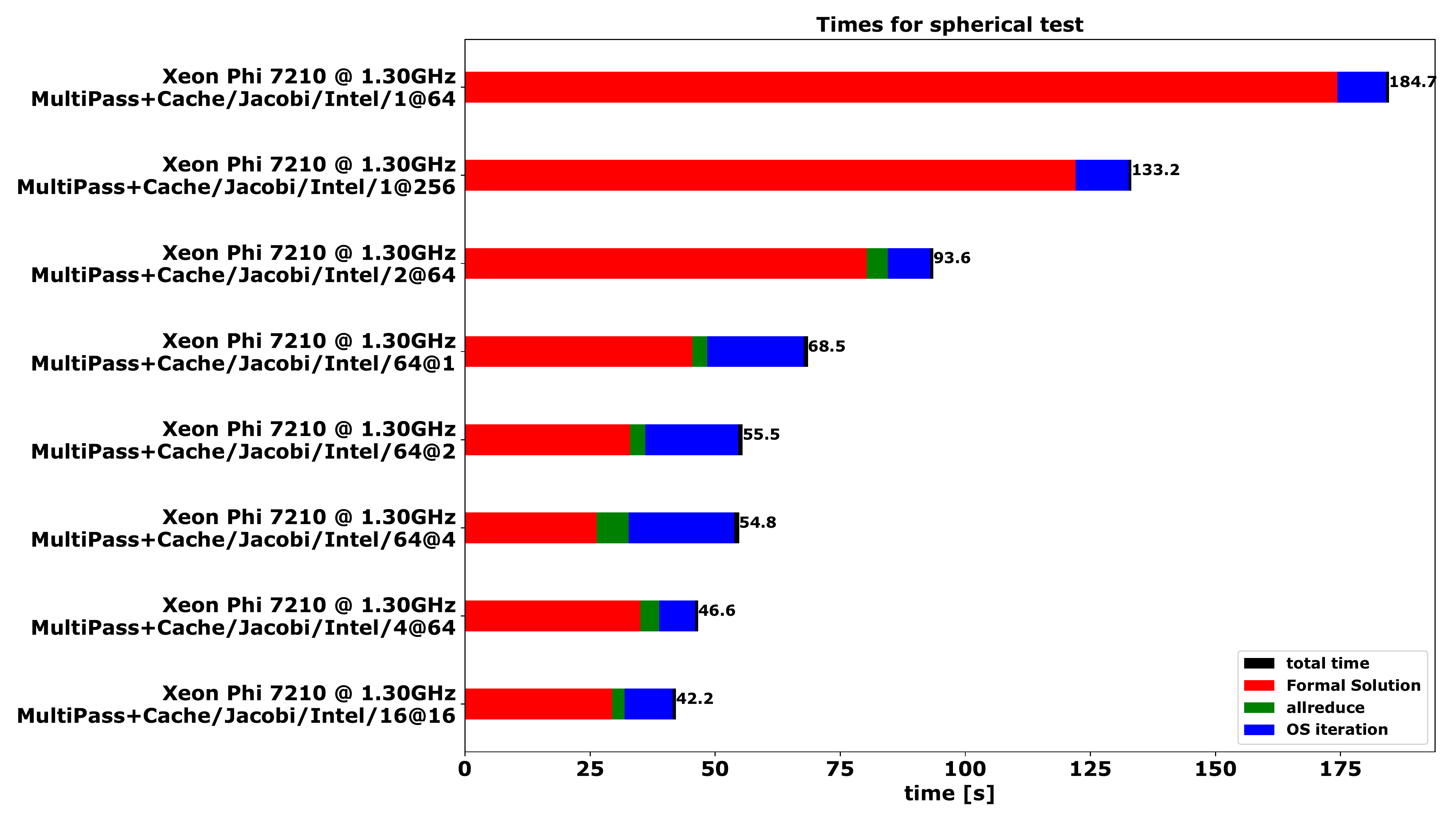}
\caption{\label{fig:KNL2}
Timing of different  MPI+OpenMP setups for the MultiPass+Cache algorithm on a many-core CPU (Xeon Phi 7210).
\expl
}
\end{figure*}

\begin{figure*}
\includegraphics[scale=0.20]{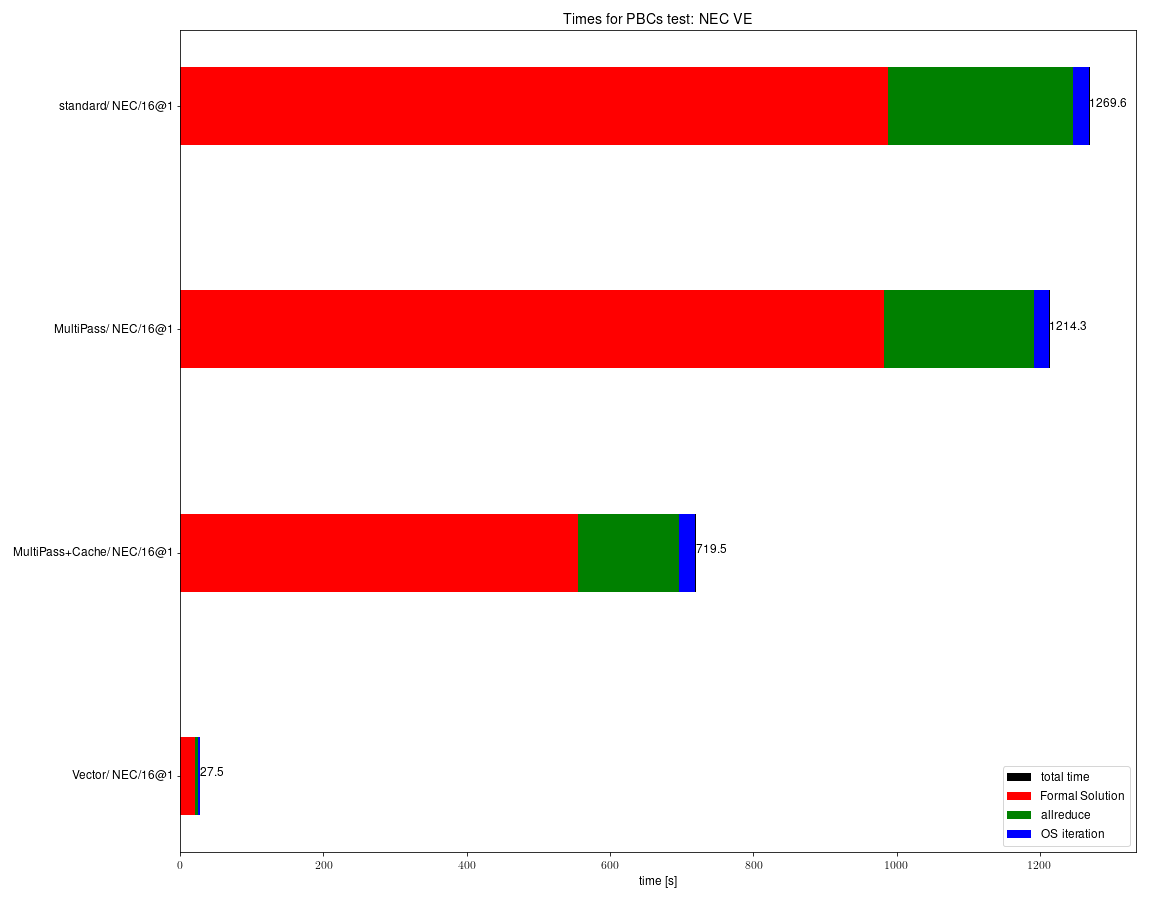}\\
\includegraphics[scale=0.20]{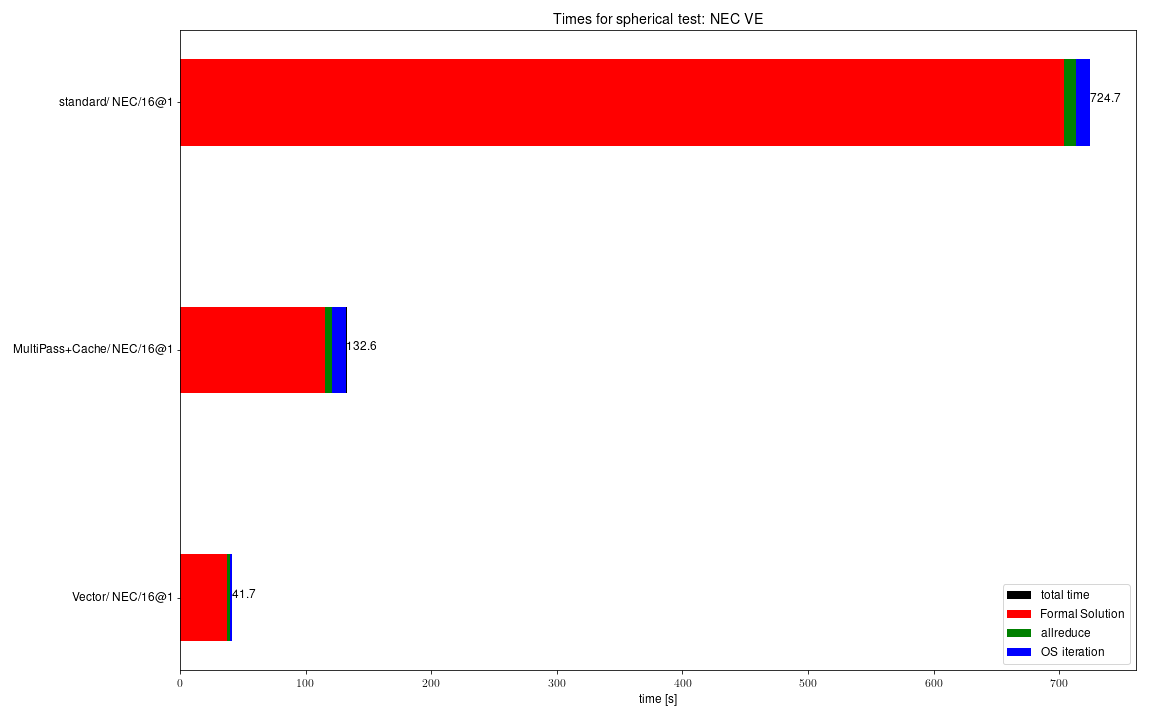}
\caption{\label{fig:NEC}
Timing of different  algorithms on a vector CPU (NEC SX-Aurora TSUBASA).
\expl
}
\end{figure*}

\begin{figure*}
\includegraphics[scale=0.25]{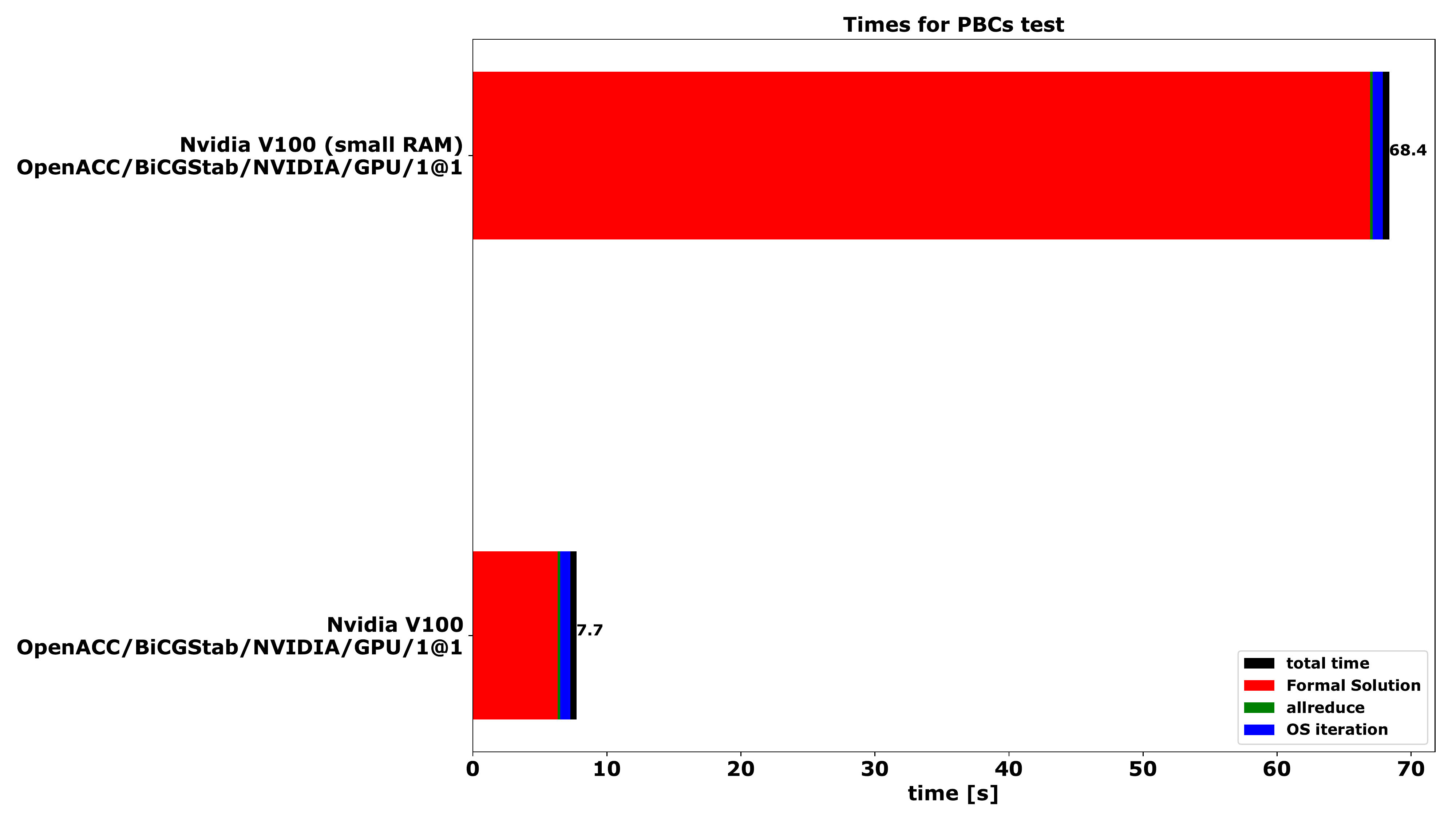}\\
\includegraphics[scale=0.25]{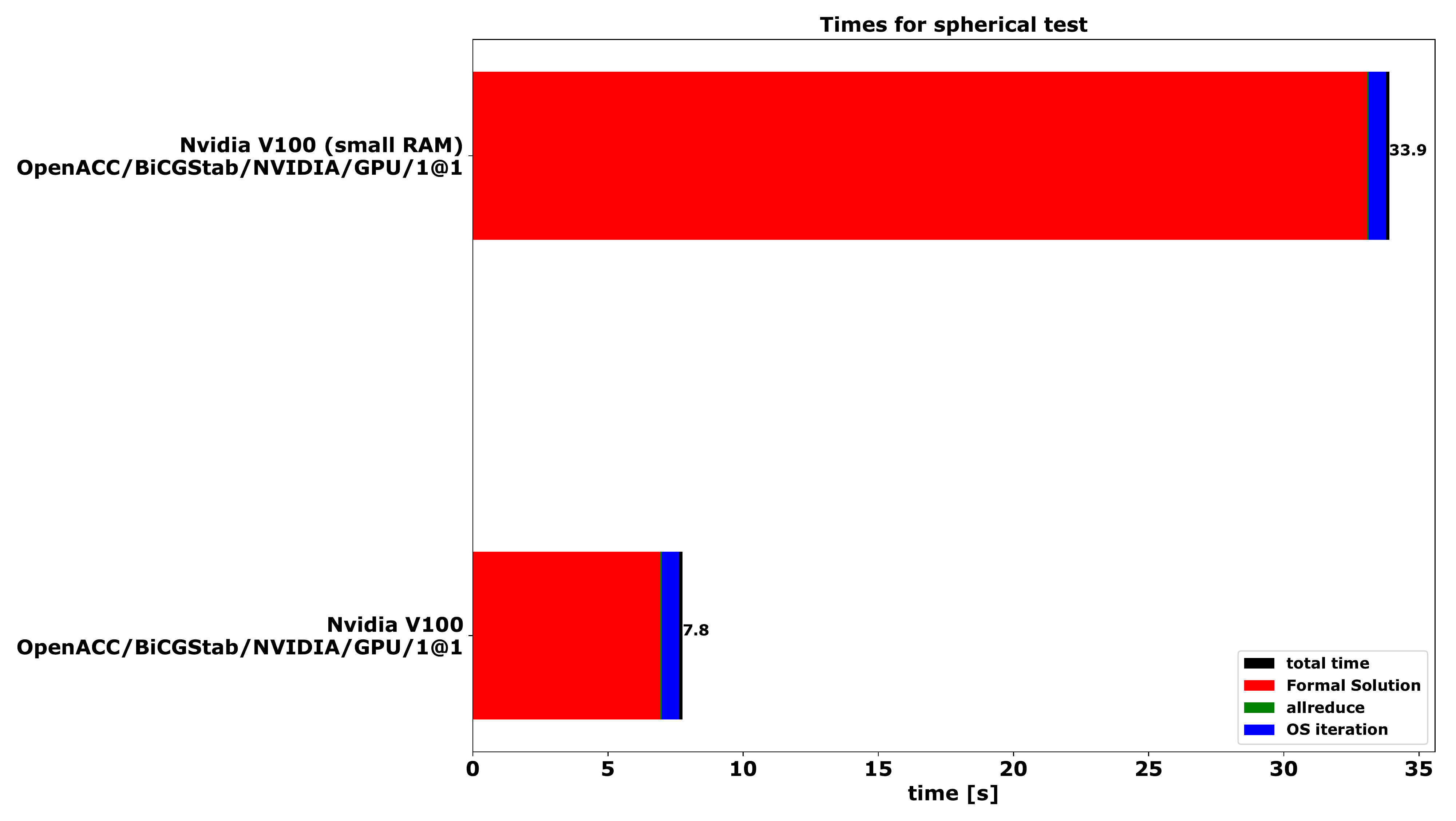}
\caption{\label{fig:PGI}
Timing of different algorithms and systems with the NVIDIA compiler with 
OpenACC support.
\expl
}
\end{figure*}

\begin{figure*}
\includegraphics[scale=0.20]{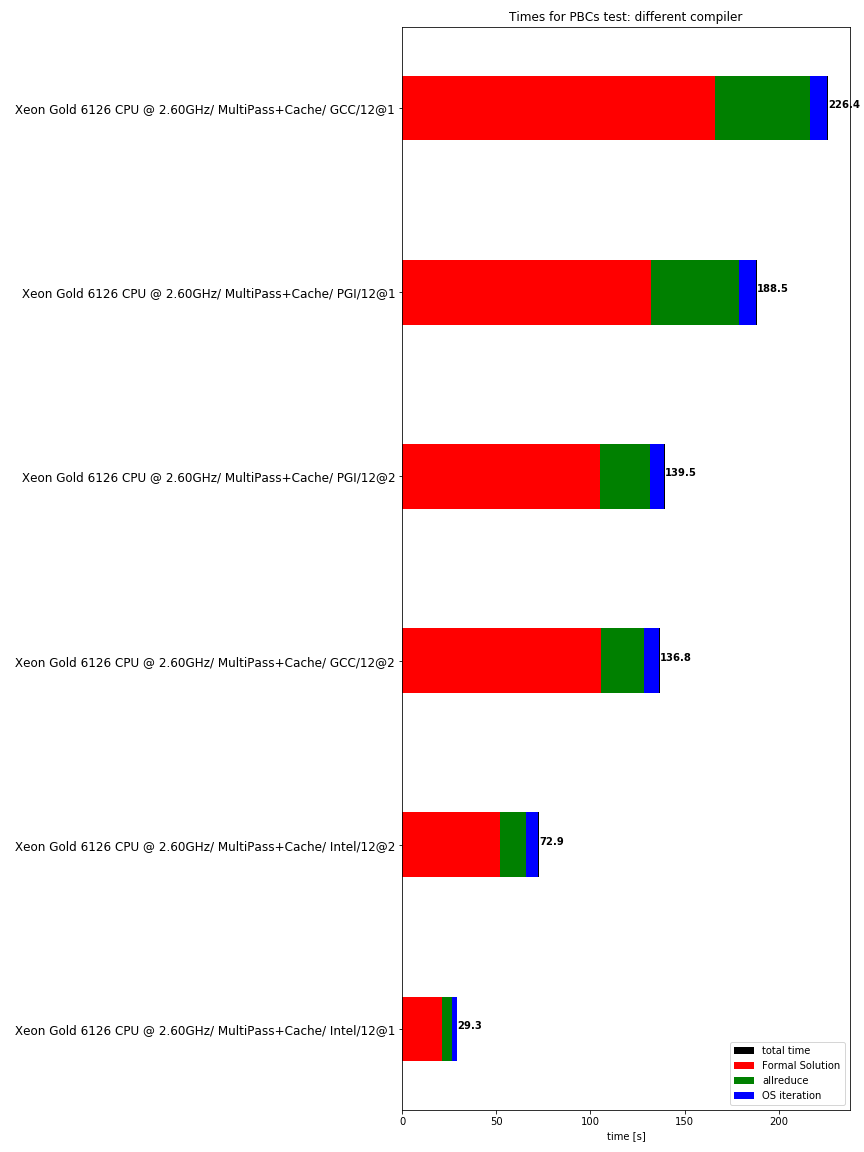}
\includegraphics[scale=0.20]{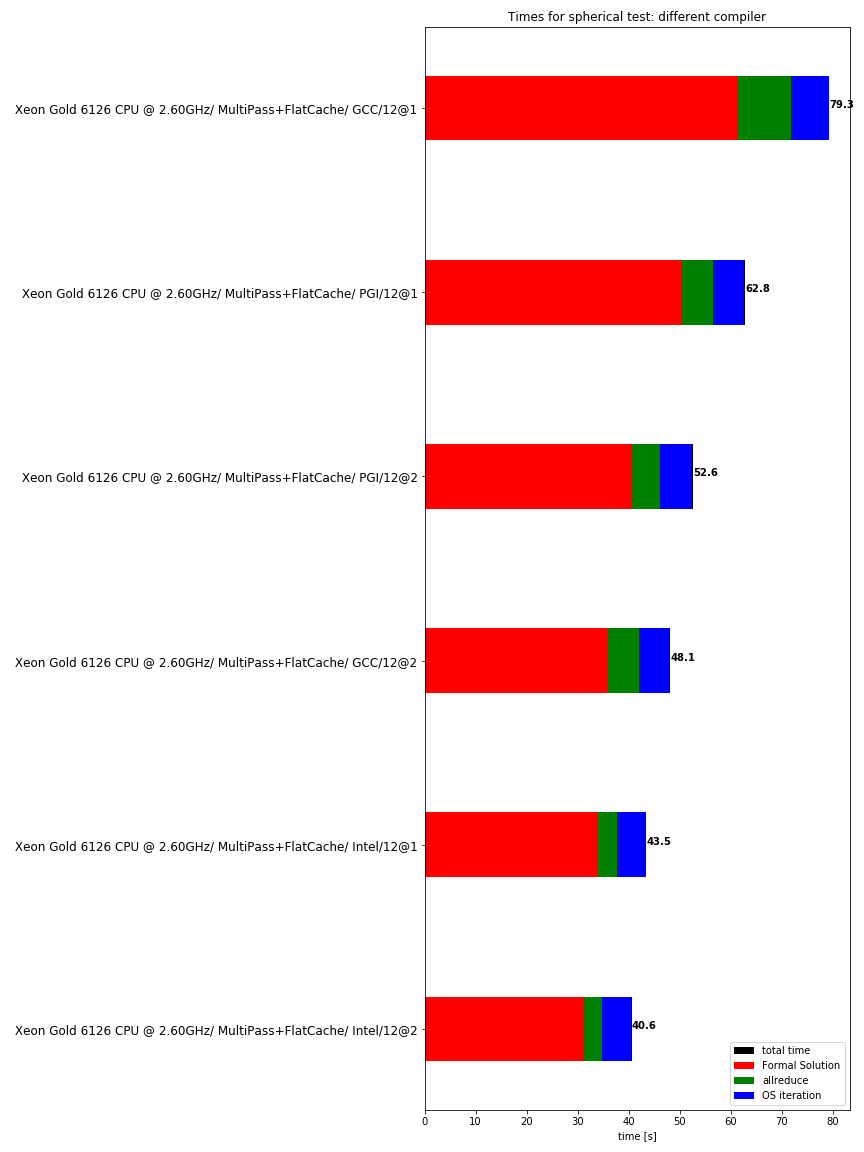}
\caption{\label{fig:compiler}
Timing for different compilers with the same algorithms for the Xeon W-3223
system.
\expl
}
\end{figure*}

\begin{figure*}
\includegraphics[scale=0.20]{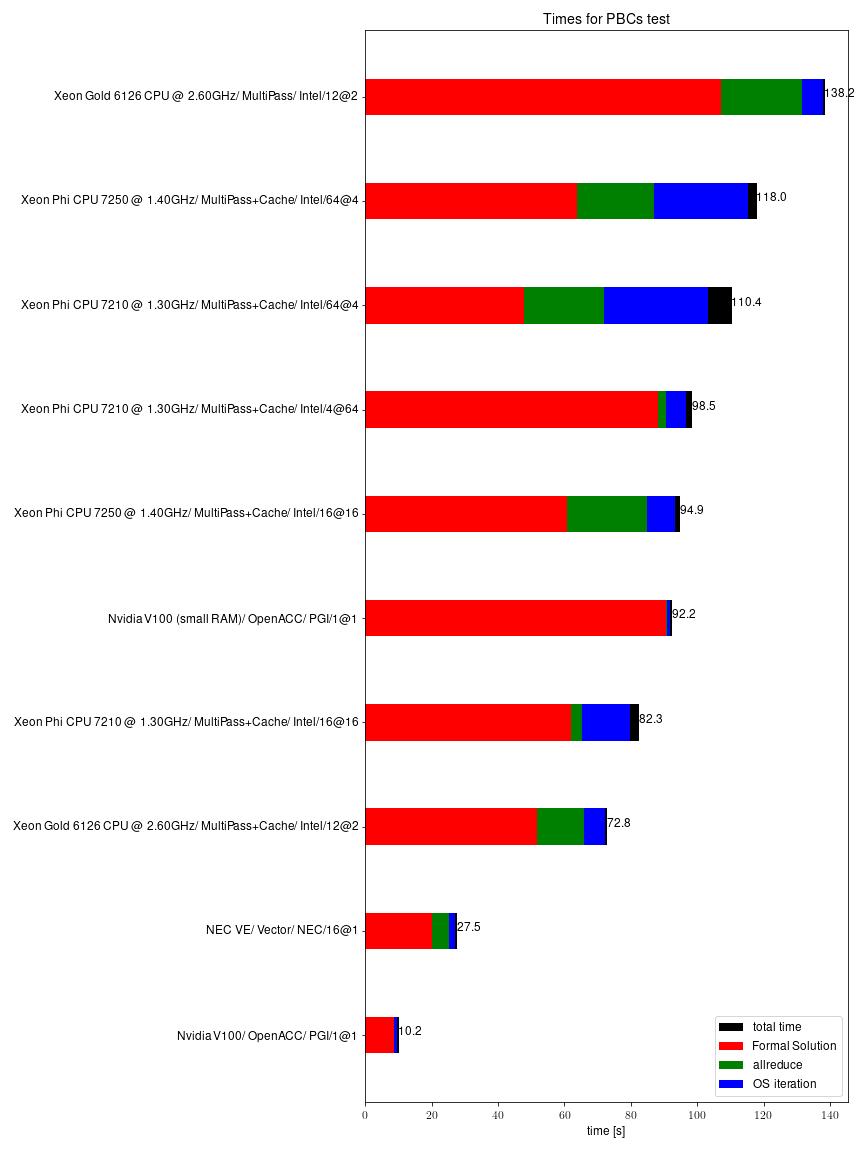}
\includegraphics[scale=0.20]{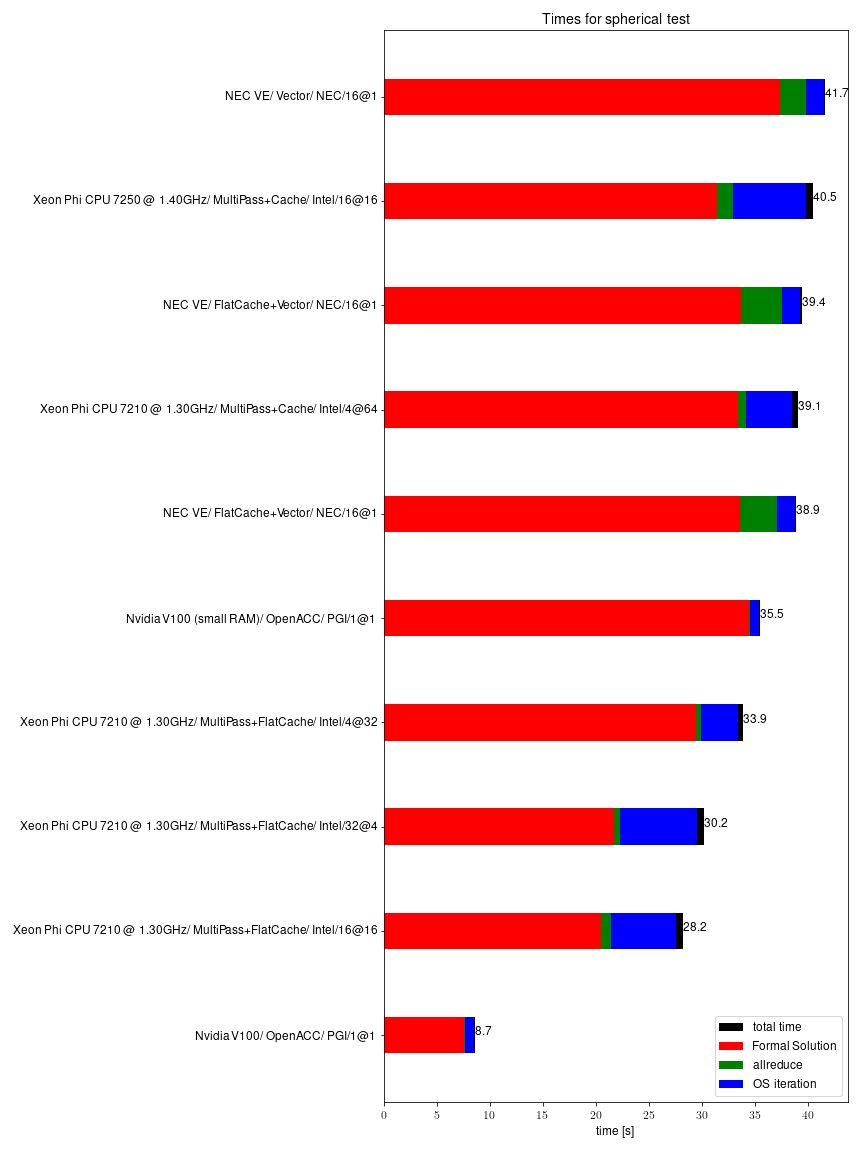}
\caption{\label{fig:total}
Timing of different algorithms for all considered
systems.
\expl.
}
\end{figure*}

\end{document}